**Non-equilibrium Magnon Engineering Enabling Significant Thermal Transport Modulation**

Takamasa Hirai*, Toshiaki Morita, Subrata Biswas, Jun Uzuhashi, Takashi Yagi, Yuichiro Yamashita, Varun Kumar Kushwaha, Fuya Makino, Rajkumar Modak, Yuya Sakuraba, Tadakatsu Ohkubo, Rulei Guo, Bin Xu, Junichiro Shiomi, Daichi Chiba, and Ken-ichi Uchida*

[1]*National Institute for Materials Science (NIMS), Tsukuba, Ibaraki, Japan.*

[2]*SANKEN, Osaka University, Ibaraki, Osaka, Japan.*

[3]*Department of Physics, Indian Institute of Technology Guwahati, Guwahati, India.*

[4]*National Institute of Advanced Industrial Science and Technology (AIST), Tsukuba, Ibaraki, Japan.*

[5]*Graduate School of Science and Technology, University of Tsukuba, Tsukuba, Ibaraki, Japan.*

[6]*Department of Mechanical Engineering, The University of Tokyo, Bunkyo, Tokyo, Japan.*

[7]*Center for Spintronics Research Network, Osaka University, Toyonaka, Osaka, Japan.*

[8]*Division of Spintronics Research Network, Institute for Open and Transdisciplinary Research Initiatives, Osaka University, Suita, Osaka, Japan.*

[9]*International Center for Synchrotron Radiation Innovation Smart, Tohoku University, Sendai, Japan.*

E-mail: HIRAI.Takamasa@nims.go.jp; UCHIDA.Kenichi@nims.go.jp

**Abstract**

Thermal conductivity, a fundamental parameter characterizing thermal transport in solids, is typically determined by electron and phonon transport. Although electrical transport properties are material-specific, recent advance in understanding phonon transport has led to new insights on controlling the thermal conductivity via phonon engineering techniques. Here, an unconventional way of artificially modulating the heat conduction in solids is demonstrated via engineering of magnon transport. Time-domain thermoreflectance measurements in ferromagnetic metal/insulator junction systems reveal that the thermal conductivity of the ferromagnetic metals and interfacial thermal conductance vary substantially depending on the spatial distribution of non-equilibrium spin currents. Systematic measurements of the thermal transport properties with changing the boundary conditions for spin currents unveil that magnons significantly modulate the heat conduction by ~10 $Wm^{-1}K^{-1}$ even in ferromagnetic metals at room temperature, upsetting the conventional wisdom that the magnon thermal conductivity is very small in metals except at low temperatures. This magnon-engineered thermal transport offers a new principle and manner for active thermal management.

## 1. Introduction

Developments in understanding and controlling thermal transport are critical for thermal management in densely packed, high performance electronic devices.[1,2] A material showing high thermal conductivity, $\kappa$, is desirable for heat dissipation and exchange, while a material showing low $\kappa$ for thermal insulation and thermoelectric conversion. However, as device dimensions shrinks, materials' $\kappa$ alters, deviating from macroscale behavior, due to more pronounced scattering processes of multiple heat carriers, e.g., electrons and phonons, at micro/nanoscale. The presence of multiple heat carriers imposes another thermal transport problem: interfacial thermal resistance, $R$, that leads to large bottlenecks to heat flow.[3] In metal/insulator junctions, commonly installed in many electronic devices, since the heat conduction in metals (insulators) is typically dominated by electrons (phonons), the large $R$ inhibits the heat flow across the heterojunction, accompanying with a finite temperature drop at the interface via electron-phonon coupling, even when using metals showing high $\kappa$.

To fully realize the multiscale thermal management depending on intended applications, technologies tailoring both $\kappa$ and interfacial thermal conductance $G$ (= $1/R$) by considering mean free paths (MFPs) of different heat carriers need to be developed. While electron contribution to $\kappa$ is connected to electrical conductivity as per the Wiedemann-Franz law, with their MFP limited to a few nanometers, significant modulation of electron $\kappa$ is challenging without resorting to phase transitions or carrier doping.[4,5] In contrast, since MFP of phonons is much longer than that of electrons and various phonon modes with different MFPs contribute to heat conduction, the micro/nanoscale thermal transport properties can be dramatically modulated without changing electrical conductivity by selectively controlling phonon scattering or phonon-phonon and electron-phonon interactions.[6,7] Various strategies for phonon-engineered thermal transport have been proposed so far, e.g., reducing $\kappa$ and/or $G$ by tuning nano-structural properties[8–11] and improving $\kappa$ and/or $G$ through electrically modulating chemical bonding,[12] lattice strain,[13] and electrical polarization.[14]

Recently, advancements in the field of spin caloritronics have provided novel thermal management principles and functionalities by introducing the spin degree of freedom into thermal transport and thermoelectric conversion.[15–17] While much focus has been on magneto-thermoelectric and thermo-spin conversion in spin caloritronics,[17–20] several studies have reported the manipulation of $\kappa$ in magnetic metal films or multilayers through magnetic field application.[21–25] Considering heat conduction dominated by electrons in metals, the magnetic-field-induced $\kappa$ change is attributed to the thermal analog of magnetoresistance effects, i.e., the change in electrical conductivity depending on a magnetization configuration. Since phonon thermal conductivity is magnetic-field- or spin-independent, the change ratio of $\kappa$ should be comparable to or less than that of the electrical conductivity as per the Wiedemann-Franz law and spin-dependent electron transport theories.[22,26] Nevertheless, a few reports show a change in $\kappa$ enough to breakdown the Wiedemann-Franz law at and above room temperature,[24,25] hinting at additional magnetic-field- or spin-dependent heat conduction mechanisms beyond conventional electron contribution. A potential mechanism is the contribution of magnons, quanta of collective motion of magnetic moments, to the heat conduction; magnons are known to carry not only spin angular momentum but also heat energy,[27] and magnon-driven $\kappa$ change in a magnetic insulator (MI) was theoretically predicted.[28] Although the recent theoretical simulation predicts that magnons contribute to heat conduction even in a ferromagnetic metal (FM) at room temperature,[29] the direct experimental observation of magnon thermal conductivity in FM has been proved only by freezing out the magnon excitation with high magnetic fields at very low temperature (< 5 K).[30,31]

In this study, we demonstrate that heat conduction in FM thin films and at FM/insulator interfaces can be significantly modulated by tuning the spatial distribution of non-equilibrium spin currents even at room temperature (Figure 1a,b). Inspired by the spin transport theory on magnetic metal/insulator junctions, we design the experimental system to elucidate spin-current-induced thermal transport engineering. Systematic investigations of thermal transport properties for various materials using an ultrafast optical pump-probe technique obtain the surprising evidence that magnons carry substantial heat even at room temperature in FM, enabling the engineering of not only $\kappa$ of FM, $\kappa_{FM}$, but also $G$ at the FM/insulator interface. Our experiment reports a groundbreaking approach in understanding heat conduction by magnon at various temperatures, even under low magnetic fields. The functionality of the magnon-engineered thermal transport paves the way for thermal engineering research and promotes spintronic thermal management[32] for applications.

## 2. Results and Discussions

### 2.1. Phenomenological prediction of thermal transport engineering by non-equilibrium spin current

To discuss the modulation of heat conduction by spin currents, we examine a transport of thermally excited spin currents in two types of FM/insulator junctions: FM/non-magnetic insulator (NI) and FM/MI junctions. When a temperature gradient is applied across the FM layer and the junction interface, a non-equilibrium spin current is excited in the FM layer. In FM/NI junctions, the spin current flowing in the thickness direction must dissipate at the interface because the spin current is not injected into NI. In contrast, in FM/MI junctions, the spin current can transmit to the MI interface via the conversion of the spin current in FM into the magnon spin current in MI despite the absence of conduction electrons in MI.[33] The difference in the boundary conditions for spin currents can induce different spin-current and spin-accumulation distributions following the spin diffusion equation.[34–36] Figure 2 presents schematics of the spatial profiles of the normalized spin current density $j_s^n$ and corresponding spin chemical potential $\mu_s^n$ in FM/NI and FM/MI junction systems. This assumes a one-dimensional system along the thickness direction with continuous spin currents at the FM/MI interface. Such a spin transport is governed by a spin diffusion length, a characteristic length in which spin currents and spin accumulation persist. As shown in Figure 2a,b, the total amount of spin currents in the FM layer adjacent to MI becomes greater than that on NI, attributed to the disappearance of spin currents at the FM/NI interface. If these spin currents convey heat, $\kappa_{FM}$ on MI should be larger than that on NI. However, when the spin diffusion length of FM $\lambda_{FM}$ is much smaller than the thickness of FM $t_{FM}$, this effect should be minimal. In contrast, when $\lambda_{FM}$ is larger than $t_{FM}$, a noticeable difference in $\kappa_{FM}$ between FM/MI and FM/NI systems emerges, owing to the stark reduction in the effective population of spins in the FM layer attached to NI (Figure 2c,d). Following this scenario, the magnitude of $G$ at the FM/insulator interface can also be affected by the type of attached insulator as it alters the transmission of spin currents and the resultant heat transfer across the interface. Although the above-mentioned spin currents in FM may comprise both conduction-electron and magnon spin currents, their $\lambda_{FM}$ values are quite different; typically, $\lambda_{FM}$ for the conduction-electron (magnon) spin current is a few nm (more than 100 nm).[37–41] Consequently, the change in $\kappa_{FM}$ of 100-nm-order thick FM films depending on the spin-current distribution is expected to be minor for conduction-electrons' spins but significant for magnons, reflecting their dominant contribution to that thermal transport.

**2.2. Experimental details of time-domain thermoreflectance**

For the proof-of-concept demonstration of the spin-current-induced thermal transport engineering, we measured the cross-plane $\kappa_{FM}$ and $G$ at the FM/NI and FM/MI interfaces by means of the time-domain thermoreflectance (TDTR) method, an optical pump-probe technique, in a front-heating and front-detection configuration (Figure 3a). The irradiation of ultrafast pump laser pulses heats up the surface of a metallic transducer deposited on a target thin film and that of probe laser pulses with varying a time delay detects the transient response of surface temperature via thermoreflectance, i.e., temperature dependence of reflectivity, enabling the determination of thermal transport properties of thin films and their interfaces.[42–45] Since the TDTR measurements do not require any microfabrication processes and electrical contacts, TDTR facilitates high-throughput and reliable thermal transport measurements on a single wafer with a wedge structure by merely shifting the position of laser spots.[46,47] For systematic and quantitative investigations, we mainly measured thermal transport properties in a double-wedge structure comprising a thickness-wedged garnet substrate and composition-spread FM thin film. Here, paramagnetic $Gd_3Ga_5O_{12}$ (GGG) and ferrimagnetic $Y_3Fe_5O_{12}$ (YIG) were selected as NI and MI, respectively, because YIG is a prominent MI for spin current physics in spintronics[48,49] and spin caloritronics[50,51] and GGG serves as a standard reference insulator for YIG owing to the same crystal structure as and the close lattice constant to YIG. By preparing a thickness-wedged-YIG/GGG substrate, we compared TDTR signals of the FM/NI and FM/MI systems using a single sample (Figure 3a,b). As one of the FM layers, we used a CoFe thin film with a gradient in the Co atomic content ratio $q$ along the film-plane direction ($y$ direction) perpendicular to the YIG-thickness-gradient direction ($x$ direction) to investigate the FM composition dependence of $\kappa_{FM}$ and $G$ (Figure 3b). The top surface of the sample was covered by an Al transducer, chosen for its well-known thermoreflectance coefficient[52]. We found that the layered structure of Al/CoFe/YIG was free from atomic diffusion and that the crystallinity and in-plane electrical conductivity did not depend on the attached insulators, confirming similar film quality of CoFe on both GGG and YIG (see Experimental Section and Figure S1 and S2, Supporting Information, for the details of structural and electrical transport measurements). All TDTR measurements were performed at room temperature and in an air atmosphere while applying a magnetic field (=50 mT along the film plane unless otherwise specified; also see Experimental Section) to assist the alignment of the magnetization direction. Here, the magnetization of CoFe and YIG was parallelly aligned via interfacial exchange interaction (Note S1 and Figure S3, Supporting Information). In our TDTR experiments, we measured the transient response of heat diffusion from the sample surface to the substrate since the heat diffusion length in FM induced by laser heating is much larger than $t_{FM}$ (Experimental Section and Figure S4, Supporting Information).

## 2.3. TDTR measurements in double-wedge CoFe/garnet substrate

Figure 3c presents the thermoreflectance signal $\varphi_{TR}$ for $Co_qFe_{1-q}$ with $q$ = 0.58 and $t_{FM}$ = 100 nm on GGG (gray symbols) and YIG with a thickness $t_{YIG}$ of 46 μm (green symbols) as a function of the delay time between pump and probe laser pulses. The $\varphi_{TR}$ data for CoFe/YIG lay below that for CoFe/GGG, suggesting faster heat diffusion in the CoFe/YIG junction than in the CoFe/GGG junction. In contrast, in the absence of the CoFe layer, the TDTR signals were nearly identical for the GGG and YIG regions, indicating minimal difference in thermal transport properties of GGG and YIG (Figure S5, Supporting Information). This result confirms that the observed change in thermal transport is significantly influenced by the CoFe layer and its interface. Note that the same TDTR signals were observed when applying out-of-plane magnetic field or in the absence of magnetic field (Figure S6, Supporting Information). The results at different $t_{YIG}$ values are shown in Note S2 and Figure S7, Supporting Information. The analysis of the TDTR signals based on a heat diffusion model allows estimation of the $\kappa_{FM}$ values on GGG and YIG ($\kappa_{FM}^{G}$ and $\kappa_{FM}^{Y}$, respectively), as well as the $G$ values at CoFe/GGG and CoFe/YIG ($G_{FM/GGG}$ and $G_{FM/YIG}$, respectively) (see the fitting curves shown as solid curves in Figure 3c and Experimental Section for details). Figure 3d shows the $\kappa_{FM}^{G}$ and $\kappa_{FM}^{Y}$ values for $Co_{58}Fe_{42}$, revealing that $\kappa_{FM}^{Y}$ (= 76±5 $Wm^{-1}K^{-1}$) was larger than $\kappa_{FM}^{G}$ (= 63±4 $Wm^{-1}K^{-1}$). This behavior cannot be explained only by electrons' contribution based on Wiedemann-Franz law (Figure S2 in Supporting Information) and is qualitatively consistent with our expectation (Figure 2). Surprisingly, the difference in $\kappa_{FM}$ between the YIG and GGG regions, $\Delta\kappa_{FM} = \kappa_{FM}^{Y} - \kappa_{FM}^{G}$, was comparable to changes due to the giant magneto-thermal resistance effect in spintronic multilayers[23] and the change ratio $\Delta\kappa_{FM}/\kappa_{FM}^{G}$ was estimated to be ~22%. Additionally, $G_{FM/YIG}$ was found to be several times larger than $G_{FM/GGG}$ (Figure 3e), supporting the existence of heat conduction concomitant with spin currents in the FM/MI junction. This result further corroborates the validity of our phenomenological prediction. Note that the $G_{FM/GGG}$ value (~$0.2\times10^{9}$ $Wm^{-2}K^{-1}$) was comparable to $G$ values at conventional metal/insulator interfaces[53] and that the $\kappa_{FM}^{G}$ value was independent of $t_{FM}$ (Figure S8, Supporting Information), which suggests the size effect of phonon heat conduction is negligible in the present sample.

We next focus on the TDTR measurements along the $y$ direction to investigate the CoFe composition dependence. Figure 4a shows the $q$ dependence of $\kappa_{FM}^{G}$ and $\kappa_{FM}^{Y}$ (see Figure S9, Supporting Information, for the temporal response of $\varphi_{TR}$ for various $q$ values). The $\kappa_{FM}$ value of CoFe exhibits composition-dependent behavior; it tends to be lower than that of pure Fe ($q$ = 0) and Co ($q$ = 100) when the Co content is $q < 0.3$ or $0.7 < q$, but higher in the intermediate composition range [note that a similar tendency has been observed for the electrical conductivity (Figure S2, Supporting Information) and for $\kappa_{FM}$ in a bulk $Co_{50}Fe_{50}$ alloy[24]]. About the difference between $\kappa_{FM}^{G}$ and $\kappa_{FM}^{Y}$, similar trends to the results for $Co_{58}Fe_{42}$, i.e., revealing $\kappa_{FM}^{Y} > \kappa_{FM}^{G}$, were observed in whole $q$ range, though the differences for $q \sim 0$ and 1 were less pronounced. The dome-shaped $q$ dependence of $\Delta\kappa_{FM}$ and $\Delta\kappa_{FM}/\kappa_{FM}^{G}$ is depicted in Figure 4b, peaking at ~10 $Wm^{-1}K^{-1}$ and ~20% for $0.2 < q < 0.7$ and diminishing with further increasing or decreasing $q$. We show the $q$ dependence of $G_{FM/GGG}$ and $G_{FM/YIG}$ in Figure 4c and the corresponding change $\Delta G = G_{FM/YIG} - G_{FM/GGG}$ and change ratio $\Delta G/G_{FM/GGG}$ in Figure 4d. While the $G_{FM/GGG}$ values of $0.1\text{-}0.3\times10^{9}$ $Wm^{-2}K^{-1}$ were similarly observed for all compositions, the $G_{FM/YIG}$ values were enhanced by 200-400% from those of $G_{FM/GGG}$. These results robustly sustain the experimental demonstration of substantial modulation of nanoscale and interfacial thermal transport in FM/insulator structures at room temperature by engineering the boundary condition for spin currents.

### 2.4. Conduction-electron spin current vs magnon spin current

The remaining question is which conduction-electrons or magnons dominantly contribute to spin-current-induced thermal transport engineering in the present FM/insulator systems. The difference in $\lambda_{FM}$ of conduction-electron and magnon spin currents (hereinafter referred to as $\lambda_{FM}^{e}$ and $\lambda_{FM}^{m}$, respectively) provides a useful hint for clarifying the origin. First, the $\lambda_{FM}^{e}$ values of Co, Fe, and CoFe are at most a few or several tens of nanometers (Table S1, Supporting Information), which is too small to explain the large change in $\kappa_{FM}$ in 100-nm-order thick FMs (Figure 2a,b). In addition, the values of $\lambda_{FM}^{e}$ are inconsistent with the $q$ dependence of $\Delta\kappa_{FM}/\kappa_{FM}^{G}$; the change in $\kappa_{FM}$ should be smaller for samples with smaller $\lambda_{FM}^{e}$, but the experimental results sometimes show the opposite tendency (e.g., for Fe and $Co_{60}Fe_{40}$). The small $\lambda_{FM}^{e}$ values also exclude the possibility of additional contributions coming from the injection of conduction-electron spin currents into FM from YIG due to the spin Seebeck effect (SSE)[50]. Subsequently, we focus on the magnon contribution. To discuss $\lambda_{FM}^{m}$, it is necessary to determine the energy scale of the magnons of interests. In typical MIs, such as YIG, a temperature gradient induces the flow of incoherent non-equilibrium high-energy magnons with a high magnon frequency (~THz).[54,55] MFPs of such high-energy magnons are shorter than those of coherent low-energy magnons excited by microwaves with GHz-order frequencies. Thus, the value of $\lambda_{FM}^{m}$ of high-energy magnons contributing to thermal transport in CoFe should be smaller than the magnon diffusion length estimated by microwave experiments (~5-20 μm)[40,41]; however, this fact alone still does not show whether $\lambda_{FM}^{m}$ of high-energy magnons is larger than $\lambda_{FM}^{e}$. Then, let us consider the case of SSE and its reciprocal called the spin Peltier effect (SPE)[51], in which high-energy sub-thermal magnons primarily contribute to the conversion between heat and spin currents. In SSE/SPE, the length scale of magnon transport reaches the order of several micrometers for MIs[56] and several tens of nanometers for FMs[57], which is much larger than $\lambda_{FM}^{e}$. Therefore, although $\lambda_{FM}^{m}$ of high-energy magnons for heat conduction has not been reported directly, it is reasonable that it becomes larger than $\lambda_{FM}^{e}$ in a similar manner to SSE/SPE. Considering the fact that $\lambda_{FM}^{e}$ is too small to explain our results, we can conclude that the magnons dominantly contribute that spin-current-induced thermal transport modulation.

To further verify the magnon contribution, we performed the TDTR measurements in a $Ni_{80}Fe_{20}$/wedged-YIG system as a control experiment. Given the notably small $\lambda_{FM}^{e}$ of NiFe alloys (~3-5 nm; Table S1, Supporting Information), the contribution of conduction-electron spin currents in $Ni_{80}Fe_{20}$/YIG is expected to be less than in CoFe/YIG. As shown in Figure 5a,b, the distinct change in $\varphi_{TR}$ similar to the case of CoFe was obtained in $Ni_{80}Fe_{20}$ with $t_{FM}$ = 100 nm, resulting in the $\Delta\kappa_{FM}/\kappa_{FM}^{G}$ of ~25% and $\Delta G_{FM/sub.}/G_{FM/GGG}$ of ~500% (see also Figure S10, Supporting Information, for the details of structural characterization). Importantly, by inserting a 10-nm-thick or 5-nm-thick Cu layer between the $Ni_{80}Fe_{20}$ and insulator layers, the difference in the temporal response of $\varphi_{TR}$ was visibly decreased, and the difference in $\kappa_{FM}$ and $G$ disappeared (Figure 5c,d; see Figure S11, Supporting Information, for the results of the film with the 5-nm-thick Cu insertion layer). Because of the absence of magnons in the Cu layer, the disappearance of the substrate dependence of $\kappa_{FM}$ and $G$ supports our interpretation that the spin-current-induced thermal transport engineering is of magnon origin.

Finally, we discuss the magnitude of $\Delta\kappa_{FM}$. Assuming that the $\Delta\kappa_{FM}$ values (~10 $Wm^{-1}K^{-1}$) observed here originate from the modulation of the magnon thermal conductivity of FM, one can see that this change is quite large because it surpasses $\kappa$ of YIG, a proficient magnon conductor. This leads to additional possibilities of magnon-related $\kappa$ modulation mechanisms, e.g., magnon-electron and magnon-phonon interactions, akin to the magnon-drag effect in thermoelectric conversion.[58,59] For further clarification of microscopic mechanisms of thermal transport engineering, including the obvious composition dependence of $\Delta\kappa_{FM}$ and $\Delta G$, the low-temperature TDTR measurements, which can easily suppress thermal magnon excitation with a strong magnetic field, would be insightful; such experiments should reveal the energy and length scales of magnons instrumental in thermal transport. Nevertheless, our measurement method uniquely allows for the assessment of magnon-driven thermal transport properties at room temperature, making it pertinent for investigating the applicability of magnon engineering in thermal management technologies.

## 3. Conclusion

In summary, we experimentally demonstrated the engineering of nanoscale thermal transport properties in FM/insulator junction systems at room temperature by controlling the boundary condition for spins. Using the TDTR technique, we found that the interfacial transmission of thermally excited spin currents not only enhanced $\kappa_{FM}$ by several tens of percent but also significantly improved $G$ compared to conventional non-magnetic metal/insulator interfaces. The observed $\kappa_{FM}$ and $G$ engineering shows the significant contribution of magnons in the heat conduction even at room temperature. The concept demonstrated in this study has broader application beyond a single interface; it could be extended to multilayers and nanocomposites, enabling the thermal transport controlling through magnon engineering even in macroscale materials. Our findings will thus provide new engineering opportunities for active thermal management based on spintronics.

## 4. Experimental Section

*Preparation of thickness-wedged-YIG/GGG substrate*: Single-crystalline YIG (111) with a thickness of 75 μm was grown on a single-crystalline GGG (111) substrate with a thickness of 0.5 mm using a liquid epitaxy method, where a portion of Y in YIG was substituted with Bi to improve lattice matching between GGG and YIG; the actual composition of YIG was $Bi_{0.1}Y_{2.9}Fe_5O_{12}$. The thickness-wedged-YIG/GGG substrates were prepared by obliquely polishing uniform YIG/GGG substrates with SiC sandpapers and alumina slurry. The range of the YIG thickness after polishing was estimated to be 0-50 μm (Figure 3b).

*Thin film deposition*: The thin films were deposited on the wedged-YIG/GGG substrate using a magnetron sputtering system (CMS-A6250X2, Comet, Inc.) at an Ar gas pressure of 0.6 Pa and room temperature. Before the deposition, the substrates were flushed at 500°C for 30 minutes to clean the surface. The $Co_qFe_{1-q}$ composition-spread films were prepared by means of a layer-by-layer wedge-shaped deposition technique,[47,60] consisting the repetition of the following three steps: (1) deposition of a wedge-shaped Co layer over a length of 7.0 mm orthogonal to the YIG wedge using a linear moving shutter, (2) rotation of the substrate by 180° in the in-plane direction, and (3) deposition of a wedge-shape Fe layer on the same area of (1), where the thickness of the $Co_qFe_{1-q}$ films after completing one (1)-(3) cycle was designed to be 0.5 nm. We repeated this sequence 200 times to fabricate the 100-nm-thick $Co_qFe_{1-q}$ composition-spread film. We also prepared 80-nm-, 150-nm-, and 200-nm-thick $Co_qFe_{1-q}$ composition-spread films on GGG to investigate the size effect on $\kappa_{FM}$ of CoFe. Although the films were fabricated by layer-by-layer deposition, the $Co_qFe_{1-q}$ films were formed with uniformly mixing Co and Fe atoms (Figure S1, Supporting Information). In addition to the CoFe composition-spread films, we fabricated uniform $Ni_{80}Fe_{20}$ (100 nm) and $Ni_{80}Fe_{20}$ (100 nm)/Cu (10 and 5 nm) films on the wedged-YIG/GGG substrate. The samples for the TDTR (electrical conductivity) measurements were covered with 48-nm-thick (5-nm-thick) Al layer without breaking the vacuum. An Al (48 nm)/wedged-YIG/GGG sample without the FM layer was also prepared specifically for TDTR measurements.

*SEM and STEM-EDS measurements*: The thickness gradient of the YIG layer was measured by cross-sectional scanning electron microscopy (SEM) observation using a focused-ion-beam (FIB)-SEM dual beam system (CarlZeiss, CrossBeam550). The structural and chemical characterization were carried out using high-angle annular dark field (HAADF) scanning transmission electron microscopy (STEM) with energy-dispersive x-ray spectroscopy (EDS) (Thermo Fisher Scientific, Titan G2 80-200 and Spectra Ultra) using the CoFe film with $t_{FM}$ = 200 nm and NiFe film with $t_{FM}$ = 100 nm. A different FIB-SEM system (Thermo Fisher Scientific, Helios G4 UX DualBeam) was used to prepare the STEM specimens by standard lift-out techniques. The STEM specimens were thinned down to 25 nm based on backscattered electrons intensity.[61] The results are displayed in Figure S1 and S10, Supporting Information.

*XRD measurement*: The crystallographic investigation was conducted by the x-ray diffraction (XRD) with Cu-Kα radiation (Rigaku, SmartLab). The x-ray was incident on the CoFe sample with $t_{FM}$ = 100 nm through a length limit slit of 0.5 mm in the direction perpendicular to the YIG thickness gradient. The irradiation center of x-ray was positioned at $(x, y)$ = (8.5 mm, 5.0 mm) for CoFe on GGG and (1.5 mm, 5.0 mm) for CoFe on YIG, as indicated in Figure 3b. The result of the XRD measurement is shown in Figure S1, Supporting Information.

*Electrical conductivity measurement*: To measure the $q$ dependence of the in-plane electrical conductivity of the CoFe film $\sigma_{FM}^{ip}$, the samples were cut out into a rectangular strip with a width of 1 mm along the direction parallel to the CoFe composition gradient from both sides of the GGG and YIG edges using a wire saw. The magnitude of $\sigma_{FM}^{ip}$ was quantified using a standard four-probe method, where the charge current of 1 mA was applied to the film through indium contacts bonded on the ends of the strip and the voltage was measured using a probe unit with a distance of 100 μm between probes. The $\sigma_{FM}^{ip}$ value for each $q$ was ascertained by adjusting the probe position along the longitudinal direction of the strip using a micrometer.

*TDTR measurement and analysis*: The schematics of the TDTR system is shown in Figure S4, Supporting Information. The system utilizes a mode-lock Er-doped fiber laser as both the pump and probe laser sources, which generates a train of pulses at a repetition rate of ~20 MHz with a pulse width of ~0.5 ps and a central wavelength at 1550 nm. The pump laser beam was modulated by a lithium niobate modulator chopping the pulses with a rectangular wave modulation at $f_{mod}$ = 200 kHz. Meanwhile, the probe laser beam was passed through a second-harmonic generator, altering its central wavelength to 775 nm. The laser power and $1/e^2$ spot diameter were set at 20 mW (< 1 mW) and 90 μm (30 μm) for the pump (probe) laser beam, respectively. Given that the diameters of the pump and probe laser spots are significantly larger than the heat diffusion length $\sqrt{D_{\mathrm{sub}}/\pi f_{\mathrm{mod}}}$ ~ 2 μm, where $D_{sub}$ is the thermal diffusivity of GGG or YIG, one-dimensional heat diffusion can be assumed in this experiment. The steady-state temperature rise at the surface of samples[42] was <1 K. The TDTR system was equipped with an electrical delay control system in which the oscillations of two lasers were synchronized to the electrical signal from a function generator, allowing for a delay between pump and probe laser pulses $\tau_d$ with 0-50 ns without requiring a mechanical delay stage.[43,45,47,62] The penetration depth of thermal diffusion in FM was roughly estimated as $\sqrt{D_{\mathrm{FM}}\tau_{\mathrm{d}}}$, with $D_{FM}$ being the thermal diffusivity of FM. Considering the $D_{FM}$ values of CoFe (NiFe) is ~$2\times10^{-5}$ $m^2s^{-1}$ (~$1\times10^{-5}$ $m^2s^{-1}$), the magnitude of $\sqrt{D_{\mathrm{FM}}\tau_{\mathrm{d}}}$ reaches approximately 1.0 μm (0.7 μm) for CoFe (NiFe) at $\tau_d$ = 50 ns, which was substantially greater than the FM thickness used in this study. To measure thermoreflectance signals, the modulated pump laser beam was irradiated onto the sample surface, i.e., the Al surface, and the probe laser beam was focused on the same spot with a specific time delay. By detecting the reflected probe laser beam using a Si adjustable balanced photoreceiver connected to a lock-in amplifier, the lock-in amplitude and phase components of the thermoreflectance signals synchronized with the modulation frequency at the time delay were detected. Through the use of both the lock-in detection technique and a resonant filter between the photodiode and the lock-in amplifier, responses to higher harmonics of the rectangular-wave-modulated pump laser pulses and dc offset were effectively removed. Note that the variation of $q$ within the laser spot size was estimated to be ~0.4 at.% by taking into account the magnitude of the composition gradient in the $Co_qFe_{1-q}$ film, which was negligibly small. During the TDTR measurements, **H** was applied along the in-plane (out-of-plane) direction using an electromagnet (by placing a permanent magnet behind the sample).

The values of $\kappa_{FM}$ and $G$ were determined through numerical simulations based on one-dimensional heat diffusion model.[63] For the analysis of TDTR data, we focused on the lock-in phase $\varphi_{TR}$, which corresponds to the ratio between the in-phase and out-of-phase components of lock-in signals.[42,64] We fitted the $\varphi_{TR}$ data at the delay time of >100 ps, where the electrons and phonons are thermalized in the Al layer. For TDTR analyses, some parameters are usually set as the fixed parameters on the heat diffusion model. In this study, the bulk values of the volumetric heat capacity $\rho C$ with $\rho$ and $C$ respectively being the density and specific heat were used for Al, Co, Fe, and Cu and the $\rho C$ value of $Co_qFe_{1-q}$ ($Ni_{80}Fe_{20}$) was calculated to be the weighted average of the bulk values of Co (Ni) and Fe by considering the composition measured by STEM-EDS. The thicknesses estimated from the STEM measurement were used for each layer in the thermal model. The thermal effusivity $e$ (= $\sqrt{\kappa\rho C} = \rho C\sqrt{D_{\mathrm{sub}}}$) of GGG and YIG was estimated to be $4.0\times10^3$ and $3.8\times10^3$ $\mathrm{Jm^{-2}s^{-0.5}K^{-1}}$, respectively, based on the TDTR experiment using the Al/wedged-YIG/GGG sample. To double-check the accuracy, the magnitude of $e$ of GGG was also verified by individually measuring $D_{\mathrm{sub}}$ using the laser flash method, $C$ using the differential scanning calorimetry, and $\rho$ using the Archimedes method for a plain GGG (111) substrate. The estimated value of $3.96\pm0.08\times10^3$ $\mathrm{Jm^{-2}s^{-0.5}K^{-1}}$ showed good consistency for GGG. The error bar for $\kappa_{FM}$ ($G$) was estimated by calculating $\varphi_{TR}$ with deviation of $D_{FM}$ and $C$ ($R$) of FM according to rectangular distribution in statistics (Figure S12, Supporting Information). We confirmed that the model has proper sensitivity to evaluate $D_{FM}$ by the sensitivity calculations (Figure S13, Supporting Information)

## Figure

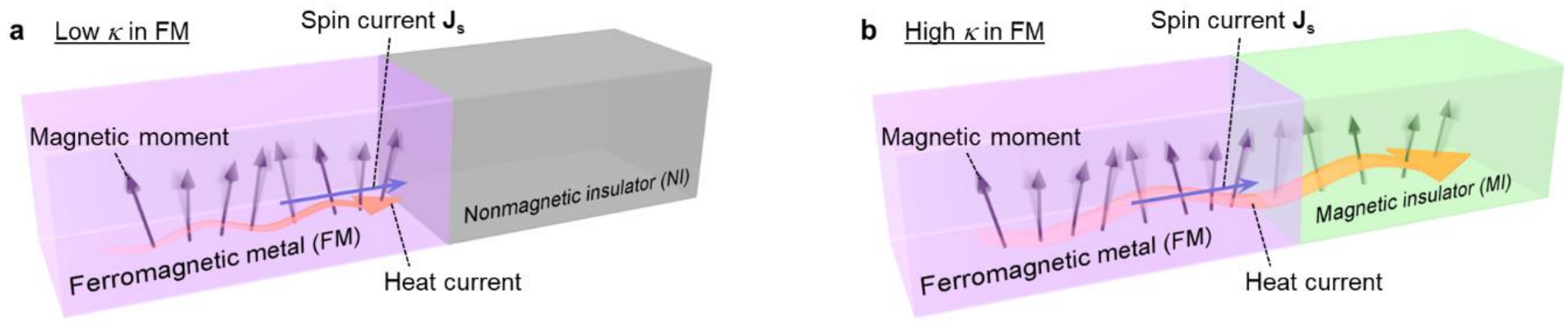

**Figure 1.** Thermal transport engineering by non-equilibrium spin currents. a,b) Schematic of magnon-engineered thermal transport in ferromagnetic metal (FM)/insulator junction structures. Thermal conductivity $\kappa$ of FM in FM/non-magnetic insulator (NI) system gets smaller than that in FM/magnetic insulator (MI) system due to disappearance (a) or transmission (b) of the spin current $\mathbf{J}_s$ at the FM/insulator interface.

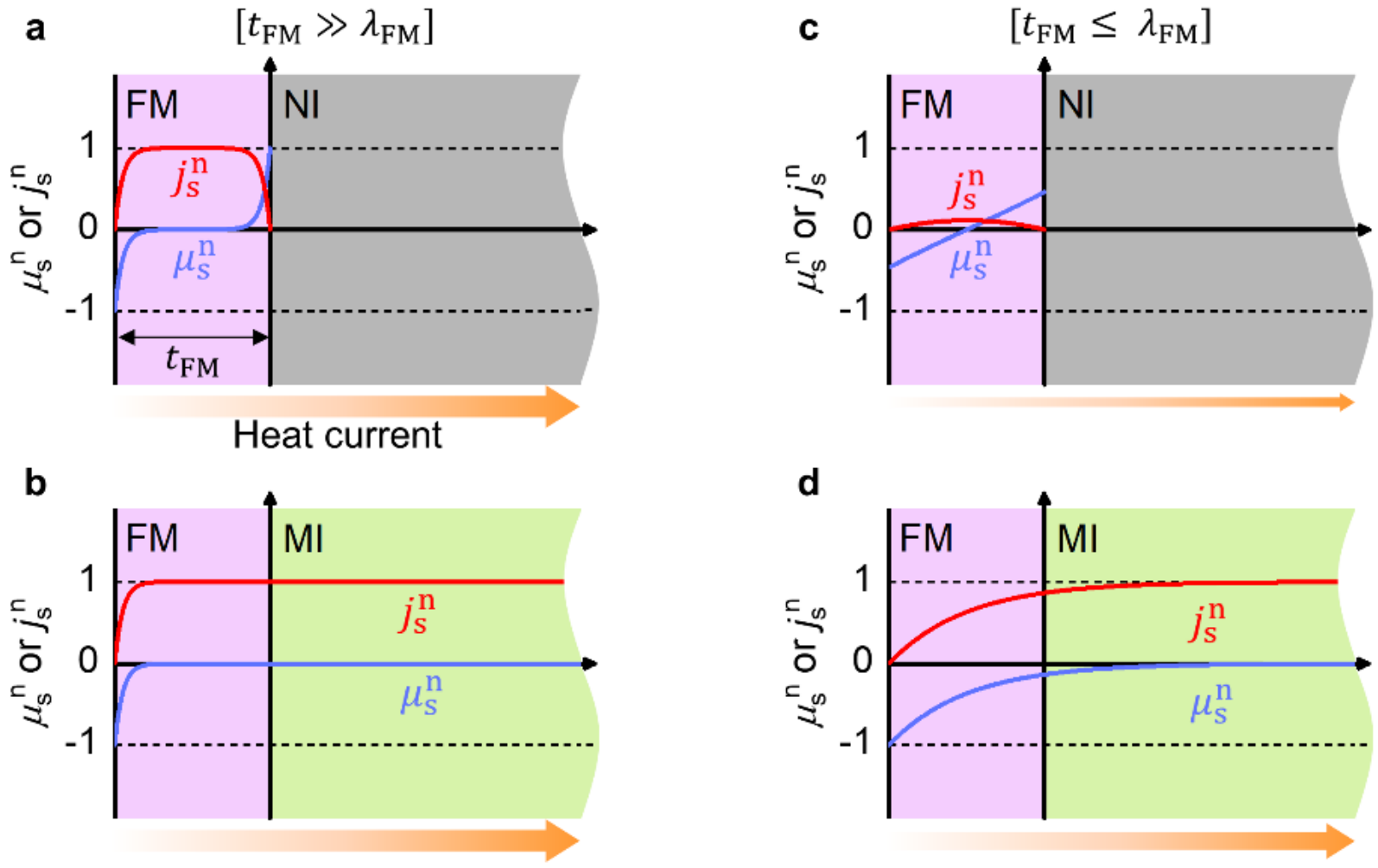

**Figure 2.** Distributions of spin current and spin chemical potential. a-d) Schematics of spatial profiles of the normalized spin current density $j_s^n$ and corresponding spin chemical potential $\mu_s^n$ following spin diffusion equation in FM/NI and FM/MI junction systems, where the spin diffusion length of the FM layer $\lambda_{FM}$ is smaller (larger) than the thickness of the FM layer $t_{FM}$ in a) and b) [c) and d)]. Orange arrows represent heat currents.

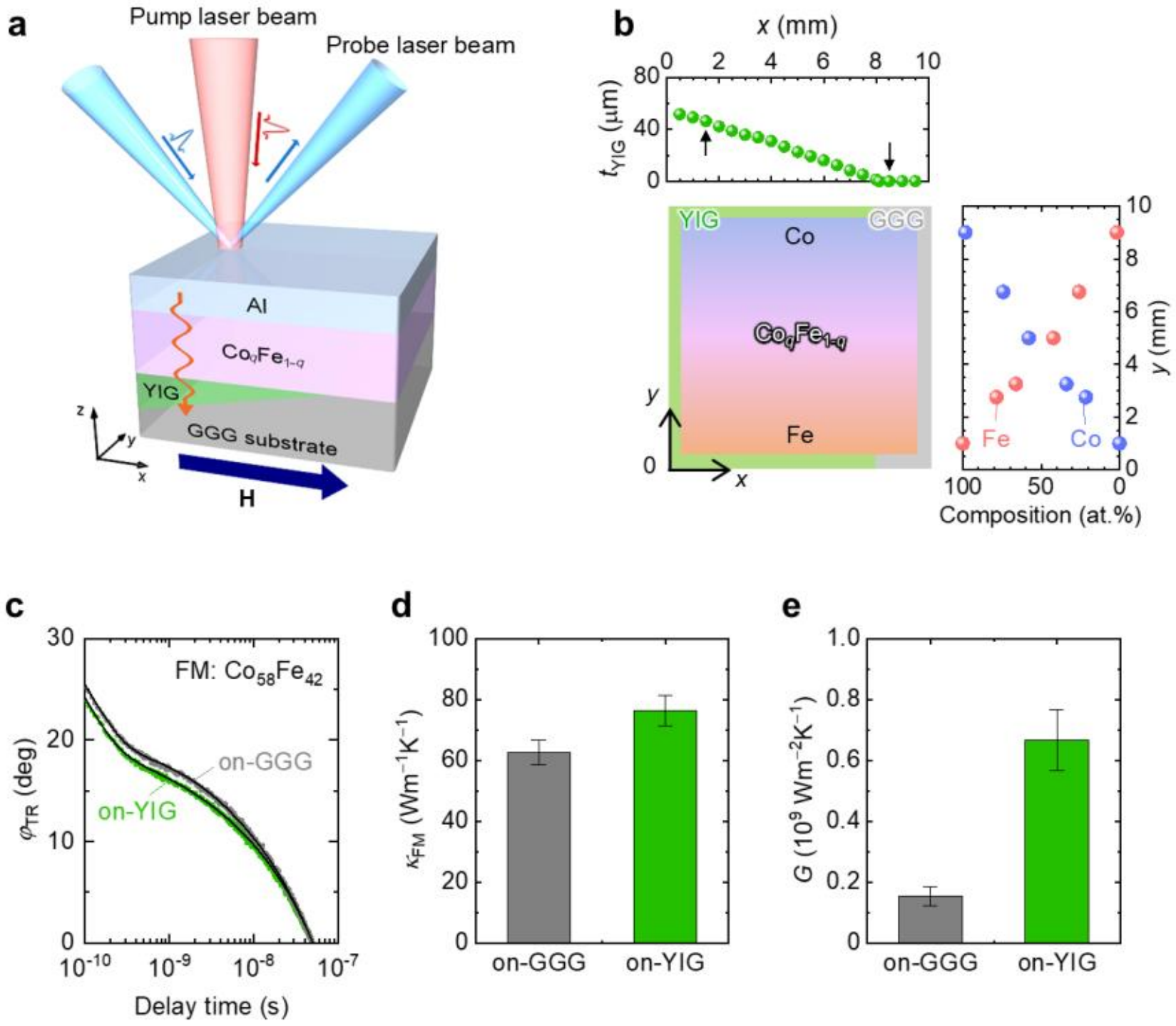

**Figure 3.** TDTR signals and thermal transport properties in CoFe/insulator junction structure. a) Schematic of the time-domain thermoreflectance (TDTR) measurement for double-wedge FM/insulator structure. Al, CoFe, $Gd_3Ga_5O_{12}$ (GGG), and $Y_3Fe_5O_{12}$ (YIG) were used as transducer, FM, NI, and MI, respectively, where the YIG thickness $t_{YIG}$ was linearly wedged along the $x$ direction and content of Co $q$ in CoFe was gradually changed from 0 to 1 along the $y$ direction. By changing the $x$ ($y$) directional position of laser spots, TDTR signals of CoFe on different insulators (with different CoFe compositions) can be obtained. During TDTR measurements, a magnetic field **H** with a magnitude of 50 mT was applied along the $x$ direction. b) Schematic of the sample from a top view and profiles of measured $t_{YIG}$ values along the $x$ direction and of Co and Fe contents along the $y$ direction (see also Experimental Section for details). Black arrows in the $t_{YIG}$ profile indicate positions at which TDTR measurements were performed for GGG ($x$ = 8.5 mm) and YIG ($x$ = 1.5 mm). c) Temporal response of $\varphi_{TR}$ for CoFe with $t_{FM}$ = 100 nm and $q$ = 0.58 ($Co_{58}Fe_{42}$) on GGG and YIG. d,e) Thermal conductivity of $Co_{58}Fe_{42}$ $\kappa_{FM}$ with $t_{FM}$ = 100 nm (d) and interfacial thermal conductance $G$ at $Co_{58}Fe_{42}$/GGG and $Co_{58}Fe_{42}$/YIG interfaces (e).

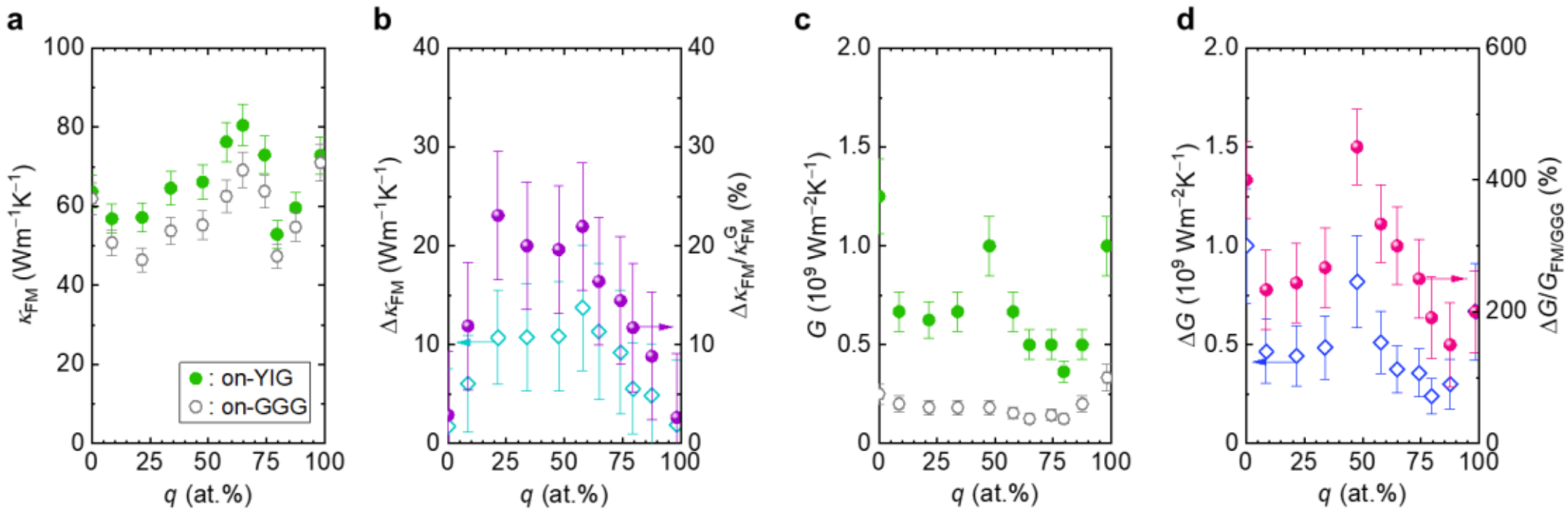

**Figure 4.** a) Temporal response of $\varphi_{TR}$ for CoFe with $q$ = 0, 0.34, 0.48, 0.74, and 0.98 on GGG and YIG. b-e) $q$ dependence of $\kappa_{FM}$ (b), change in $\kappa_{FM}$, $\Delta\kappa_{FM} = \kappa_{FM}^{Y} - \kappa_{FM}^{G}$, and change ratio, $\Delta\kappa_{FM}/\kappa_{FM}^{G}$, with $\kappa_{FM}^{G(Y)}$ being the magnitude of $\kappa_{FM}$ on GGG (YIG) (c), $G$ at FM/GGG and YIG interfaces (d), and change in $G$, $\Delta G = G_{FM/YIG} - G_{FM/GGG}$, and change ratio, $\Delta G/G_{FM/GGG}$, with $G_{FM/GGG(YIG)}$ being the magnitude of $G$ at the FM/GGG (FM/YIG) interface (e).

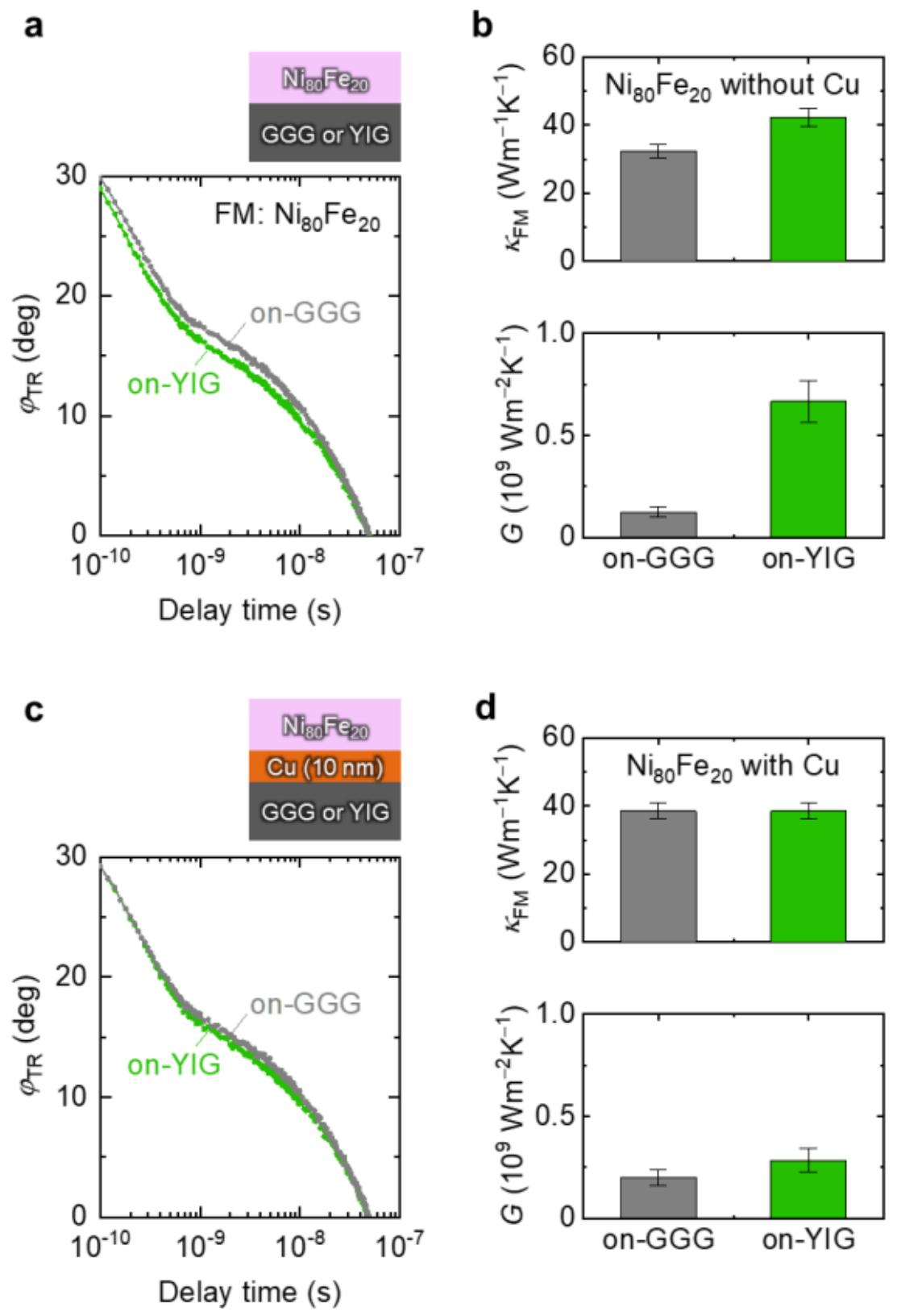

**Figure 5.** a,c) Temporal response of $\varphi_{TR}$ for $Ni_{80}Fe_{20}$ (a) and $Ni_{80}Fe_{20}$/Cu (c) films with a $Ni_{80}Fe_{20}$ (Cu) thickness of 100 nm (10 nm) on GGG and YIG. b,d) $\kappa_{FM}$ (upper panel) and $G$ (lower panel) for the $Ni_{80}Fe_{20}$ (b) and $Ni_{80}Fe_{20}$/Cu (d) films on GGG and YIG.

**Supporting Information**

Supporting Information is available from the Wiley Online Library or from the author.

**Acknowledgements**

T.H. and T.M. contributed equally to this work. The authors thank G. E. W. Bauer, R. Iguchi, T. Koyama, A. Srinivansan, and P. Alagarsamy for valuable discussions and T. Hiroto, K. Suzuki, R. Toyama, M. Isomura, A. Takahagi, K. Takamori, and R. Nagasawa for technical support. This work was partially supported by CREST "Creation of Innovative Core Technologies for Nano-enabled Thermal Management" (No. JPMJCR17I1) and ERATO "Magnetic Thermal Management Materials Project" (No. JPMJER2201) from JST, Japan; Grant-in-Aid for Research Activity Start-up (No. 22K20495) and Grant-in-Aid for Scientific Research (S) (No. 22H04965) from JSPS KAKENHI, Japan; and NEC Corporation. A part of this work was supported by the Electron Microscopy Unit, National Institute for Materials Science (NIMS). T.H. acknowledged support from the Thermal and Electric Energy Technology Foundation. T.M. was supported by Program for Leading Graduate Schools "Interactive Materials Science Cadet Program".

Supporting Information

**Non-equilibrium Magnon Engineering Enabling Significant Thermal Transport Modulation**

Takamasa Hirai*, Toshiaki Morita, Subrata Biswas, Jun Uzuhashi, Takashi Yagi, Yuichiro Yamashita, Varun Kumar Kushwaha, Fuya Makino, Rajkumar Modak, Yuya Sakuraba, Tadakatsu Ohkubo, Rulei Guo, Bin Xu, Junichiro Shiomi, Daichi Chiba, and Ken-ichi Uchida*

**Note S1. Measurement of transverse thermoelectric effects.**

To investigate magnetic properties of our ferromagnetic metal (FM)/$Y_3Fe_5O_{12}$ (YIG) interface) stacks, we performed thermoelectric voltage measurements by applying a magnetic field and temperature gradient orthogonal to each other (Figure S3a). The sample, Al(2 nm)/$Co_{50}Fe_{50}$ or $Ni_{80}Fe_{20}$(10 or 100 nm) stacked on the YIG/$Gd_3Ga_5O_{12}$ substrate of a 2 × 7 mm chip, was mounted on a brass heat sink via a sapphire substrate and thermal grease. A heater was attached to the top surface of the sample via a sapphire substrate and thermal grease. Here, a temperature difference $\Delta T$ between the heater and heat sink was controlled through monitoring each temperature using T-type thermocouples. Figure S3b-d shows the magnetic field $H$ dependence of the transverse thermoelectric voltage $V$ at various $\Delta T$ values for $Co_{50}Fe_{50}$(10 nm), $Co_{50}Fe_{50}$(100 nm), $Ni_{80}Fe_{20}$(10 nm), and $Ni_{80}Fe_{20}$(100 nm) films on YIG, respectively. These $V$-$H$ curves were arising from two effects: the anomalous Nernst effect (ANE) in the FM layer and a combination of the spin Seebeck effect (SSE) and inverse spin Hall effect (ISHE). In the latter case, the application of temperature gradient generates a spin current across the interface of FM/YIG by SSE and the spin current is converted into a charge current in the FM layer by ISHE. The single sharp change in $V$ due to the magnetization reversal in all the $V$-$H$ curves (see also Figure S3f) confirmed the magnetization of FM and YIG were parallelly aligned along the in-plane direction even at the interface via exchange interaction (note that if an additional interfacial magnetic anisotropy and a twisted spin state at the interface is introduced in FM/YIG stacks, $V$-$H$ curves exhibit complicated magnetization reversal processes[S1,S2]). We also found that the saturation value of $V$, $V_s$, defined as average of $V$ at 50 mT<$\mu_0 H$<100 mT, for 10-nm thick $Co_{50}Fe_{50}$ ($Ni_{80}Fe_{20}$) was almost same as (much larger than) that for 100-nm thick (Figure S3g,h). Considering a series thermal circuit of FM/YIG/GGG based on the Fourier's law, the temperature gradient in the FM layer is determined only by that in YIG/GGG substrate, i.e., the temperature gradient in the FM layer is independent of the FM thickness.[S2,S3] Thus, the $V_s$ value originating from ANE is independent of the FM thickness, while that from SSE-induced-ISHE monotonically increase with decreasing the FM thickness. These results indicate the tiny (large) contribution of SSE-induced-ISHE in $Co_{50}Fe_{50}$/YIG ($Ni_{80}Fe_{20}$/YIG), consistent with previous research.[S1,S2,S4]

**Note S2. Time-domain thermoreflectance measurements at different YIG thicknesses.**

Figure S7 shows the temporal response of $\varphi_{TR}$ for $Co_{58}Fe_{42}$ on GGG and YIG with various thickness $t_{YIG}$ values. The time-domain thermoreflectance (TDTR) signal at $t_{YIG}$ = 17 μm (Figure S7a) was same as that at $t_{YIG}$ = 46 μm (Figure S3c), while at thinner $t_{YIG}$, different results were obtained; $\varphi_{TR}$ got close to that for GGG at $t_{YIG}$ = 3 μm and lay below that for YIG with $t_{YIG}$ = 46 μm at $t_{YIG}$ = 1 μm. One plausible origin of different TDTR signals is the change in non-equilibrium magnon accumulation in the CoFe/YIG system because the $t_{YIG}$ value becomes comparable to or smaller than the magnon diffusion length. According to our phenomenological prediction (Figure 2), the decrease in $t_{YIG}$ should reduce the thermal conductivity of CoFe, shifting $\varphi_{TR}$ closer to $\varphi_{TR}$ for GGG. However, the decrease in $t_{YIG}$ also induces a reduction of the thermal conductivity of YIG due to the size effect, which greatly causes the $\varphi_{TR}$ signal shift away from $\varphi_{TR}$ for GGG (Figure S13). In the current experimental setup and analyses, it is difficult to distinguish between these two contributions, so a detailed investigation will be left as a task for the future.

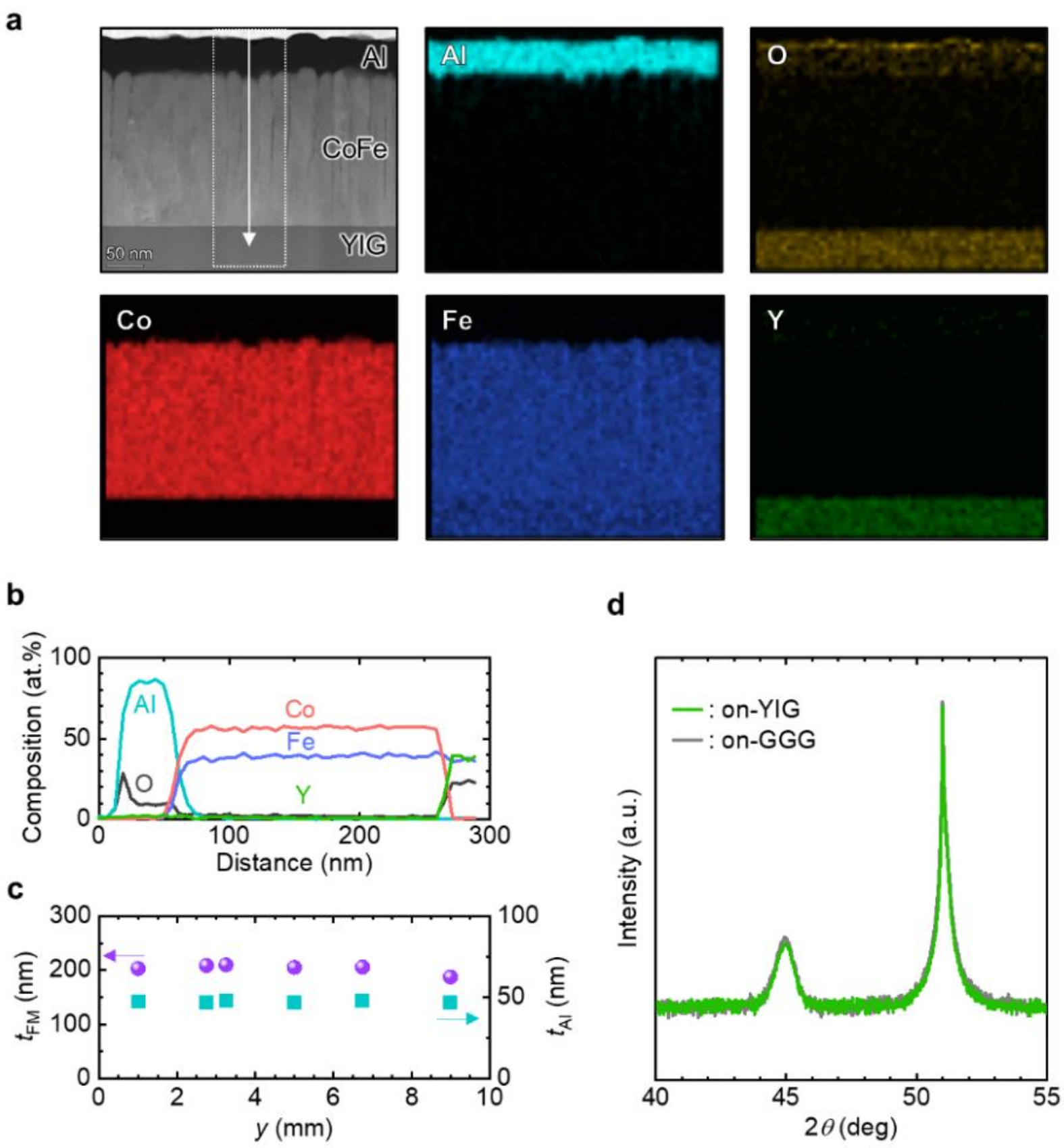

**Figure S1.** a) Cross-sectional high-angle annular dark-field scanning transmission electron microscopy (HAADF-STEM) image and corresponding elemental maps of CoFe film on a thickness-wedged-$Y_3Fe_5O_{12}$ (YIG)/$Gd_3Ga_5O_{12}$ (GGG) substrate obtained using energy-dispersive x-ray spectroscopy (EDS), where the CoFe composition in images is approximately $Co_{58}Fe_{42}$. b) EDS line profiles of constituent elements along the out-of-plane direction along the white arrow in dashed box region shown in the STEM image in a). c) Thicknesses of CoFe ($t_{FM}$) and Al ($t_{Al}$) layers for CoFe/YIG sample with $t_{FM}$ = 200 nm at each $y$ position in Fig. 2b, as determined by STEM-EDS. Almost uniform $t_{FM}$ and $t_{Al}$ values were obtained throughout the sample. d) X-ray diffraction profiles of CoFe film with $t_{FM}$ = 100 nm on GGG and YIG. In both films, the CoFe peaks with similar intensities and same peak centers were detected at around 45° in addition to GGG or YIG 444 peak around 51°, indicating that the crystal structure of CoFe does not depend on the substrate.

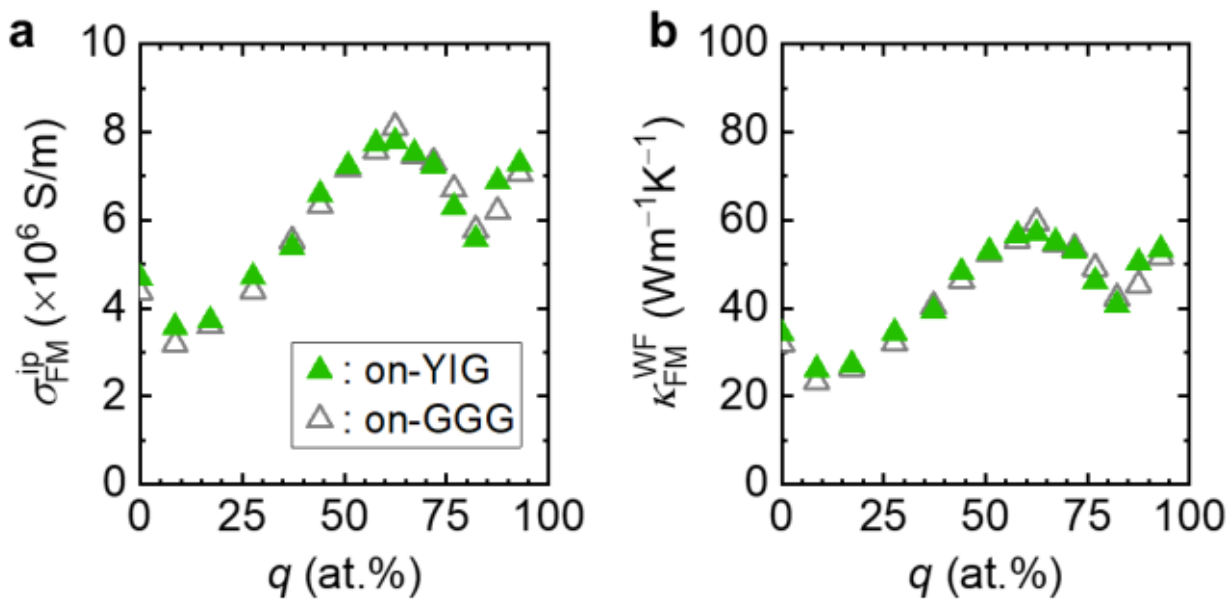

**Figure S2.** a,b) $q$ dependence of the in-plane electrical conductivity (a) and thermal conductivity calculated by the Wiedemann-Franz law (b) of CoFe on GGG and YIG. $q$ denotes the Co content at each position of the CoFe composition-spread film.

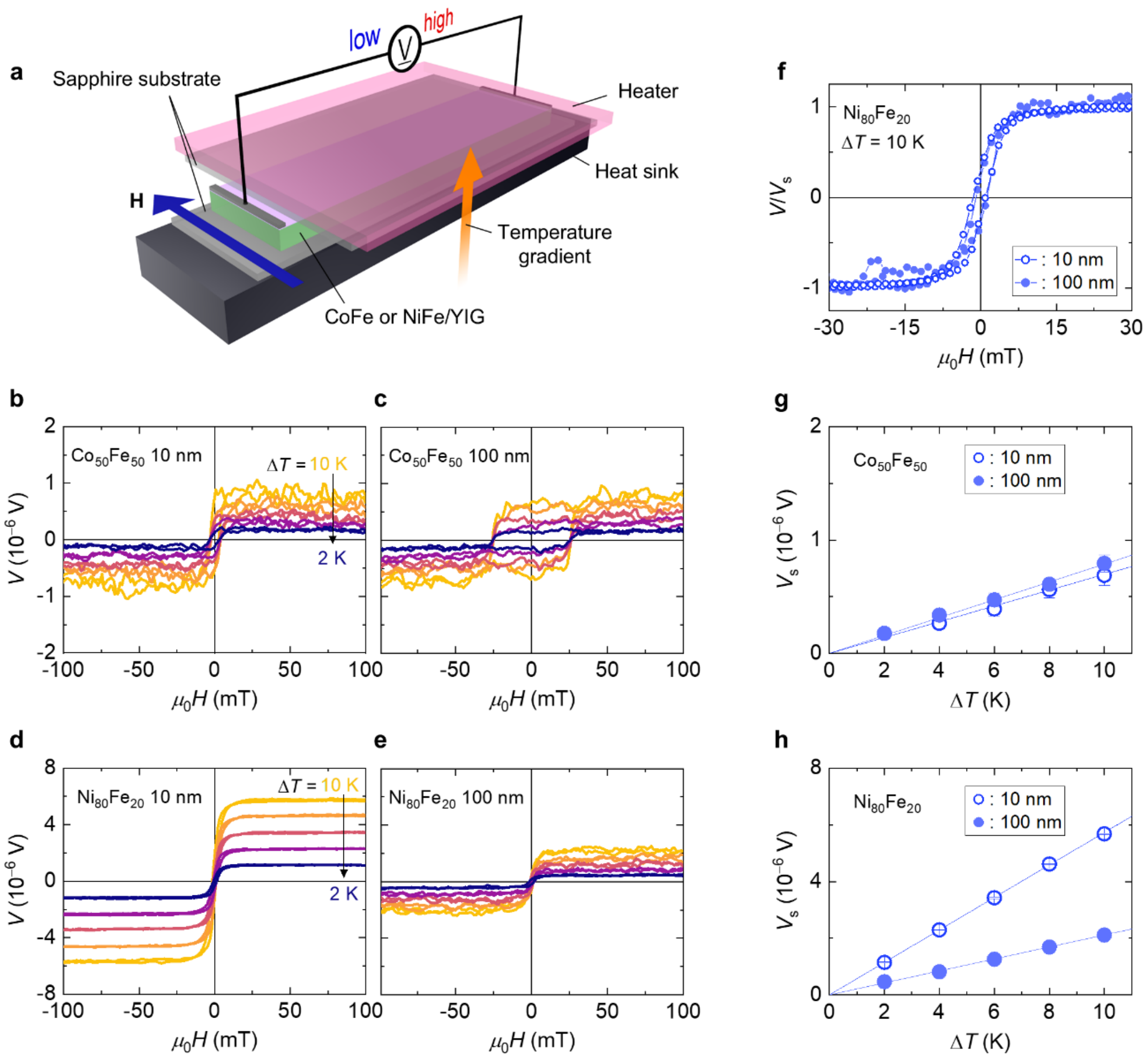

**Figure S3.** a) Schematic of the transverse thermoelectric voltage $V$ measurement with applying a magnetic field **H** with the magnitude of $H$ and temperature gradient. b,c) $H$ dependence of $V$ in the $Co_{50}Fe_{50}$ film with a thickness of 10 (b) and 100 nm (c) at various values of temperature differences between heater and heat sink $\Delta T$. d,e) $H$ dependence of $V$ in the $Ni_{80}Fe_{20}$ film with a thickness of 10 (d) and 100 nm (e) at various $\Delta T$ values. f) $H$ dependence of $V/V_s$ in the $Ni_{80}Fe_{20}$ film at $\Delta T = 10$ K, where $V_s$ is average of $V$ at 50 mT<$\mu_0 H$<100 mT with $\mu_0$ being the vacuum permeability. g,h) $\Delta T$ dependence of $V_s$ for $Co_{50}Fe_{50}$ (g) and $Ni_{80}Fe_{20}$ (h).

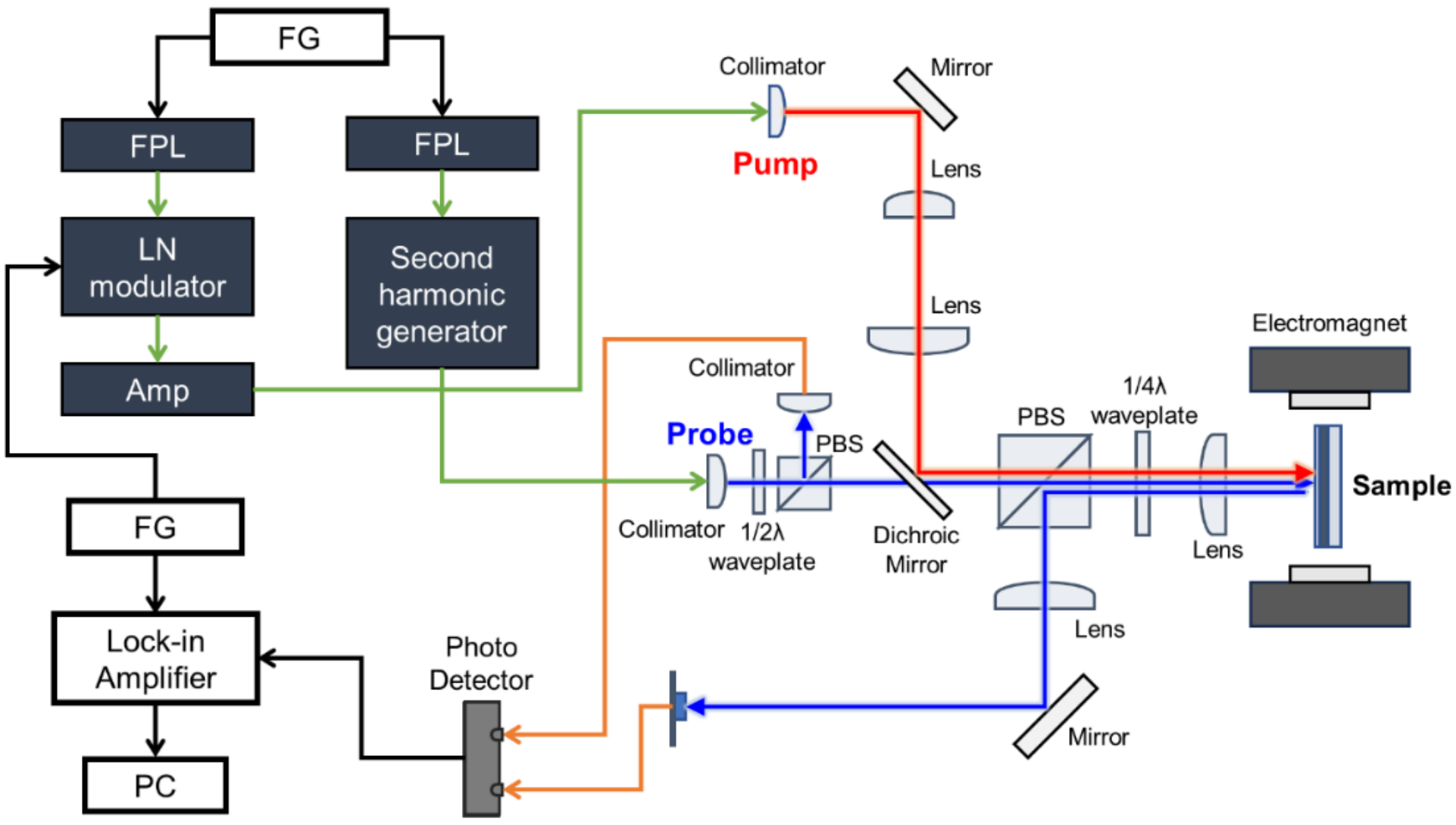

**Figure S4.** Schematic diagram of our time-domain thermoreflectance (TDTR) system. FPL is a fiber-based pulsed laser, FG a function generator, and PBS a polarizing beam splitter.

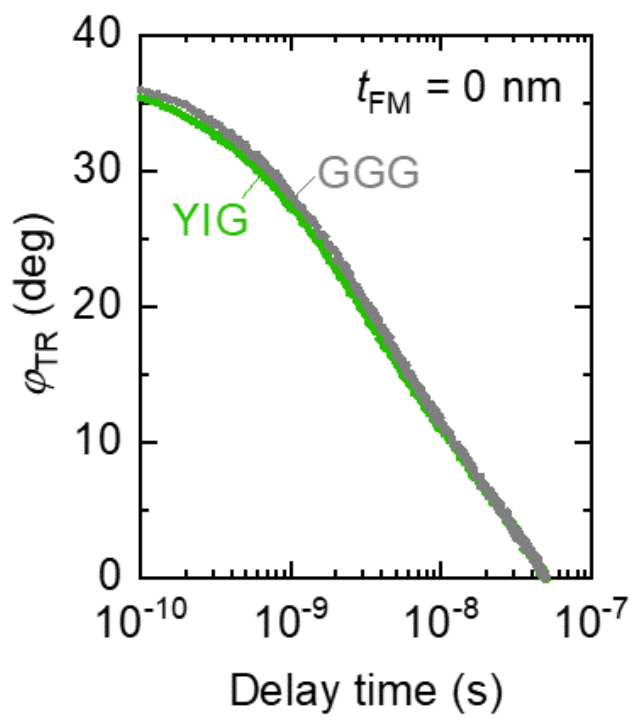

**Figure S5.** TDTR signals in the sample without the CoFe layer. Temporal response of thermoreflectance signal $\varphi_{TR}$ for the junction structure of Al/GGG and Al/YIG. The tiny difference in $\varphi_{TR}$ indicates that the trivial change in thermal parameter between GGG and YIG (see Experimental Section) did not affect the observed change between CoFe/GGG and CoFe/YIG junction structures.

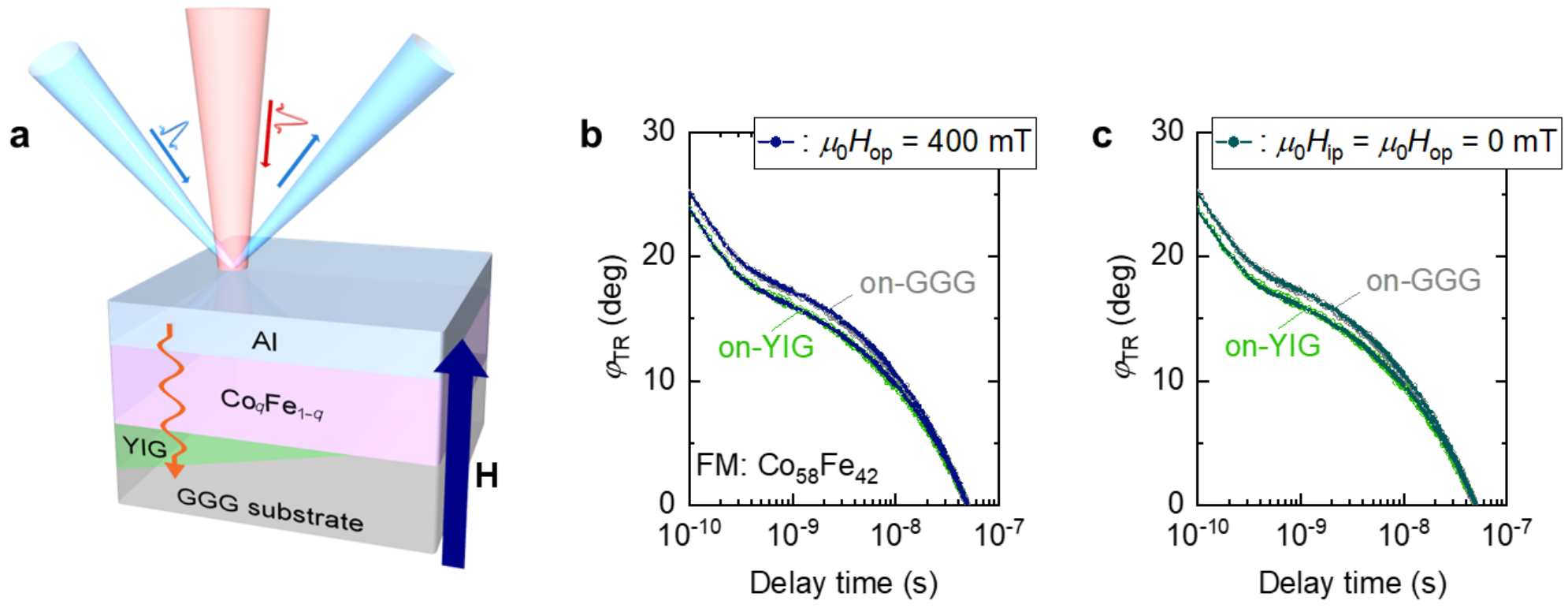

**Figure S6.** a) Schematic of the TDTR measurement with applying **H** along out-of-plane direction. b,c) Temporal response of $\varphi_{TR}$ for $Co_{58}Fe_{42}$ on GGG and YIG at $\mu_0 H_{op}$ = 400 mT (b) and $\mu_0 H_{ip}$ = $\mu_0 H_{op}$ = 0 mT (c), where $\mu_0$ and $H_{op(ip)}$ represent the vacuum permeability and magnitude of out-of-plane (in-plane) **H**, respectively. Data at $\mu_0 H_{ip}$ = 50 mT (Figure 3c) are also shown for comparison.

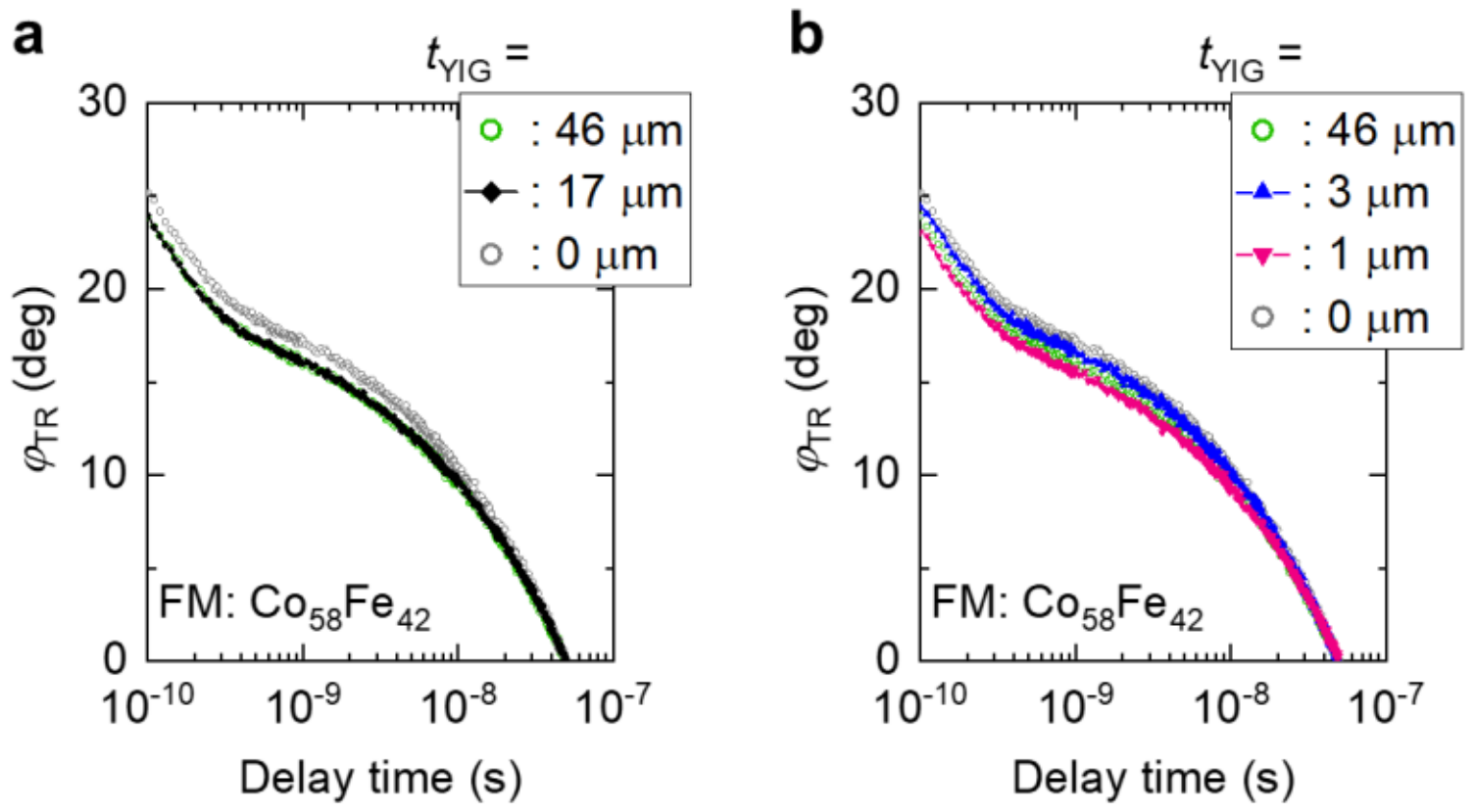

**Figure S7.** a,b) Temporal response of $\varphi_{TR}$ for $Co_{58}Fe_{42}$ on GGG and YIG with various thickness $t_{YIG}$ values; at $t_{YIG}$ = 17 μm (a) and $t_{YIG}$ = 3 and 1 μm (b). Data at $t_{YIG}$ = 46 and 0 μm (on GGG) (Figure 3c) are also shown for comparison.

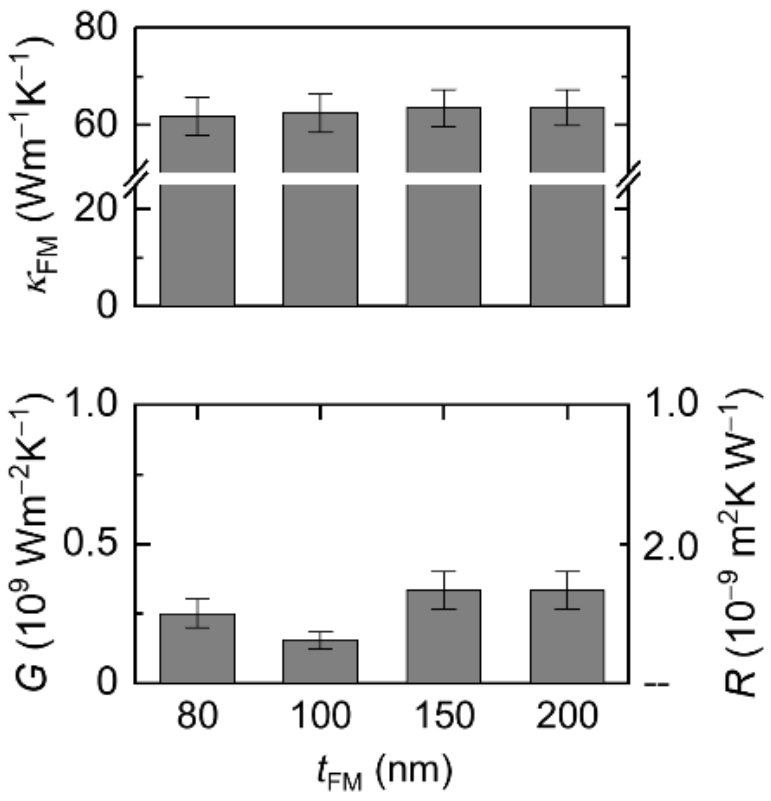

**Figure S8.** Thermal conductivity of the CoFe film $\kappa_{FM}$ on GGG (upper panel) and interfacial thermal conductance $G$ and resistance $R = 1/G$ at the CoFe/GGG interface (lower panel) for $Co_{58}Fe_{42}$ with $t_{FM}$ = 80, 100, 150, and 200 nm.

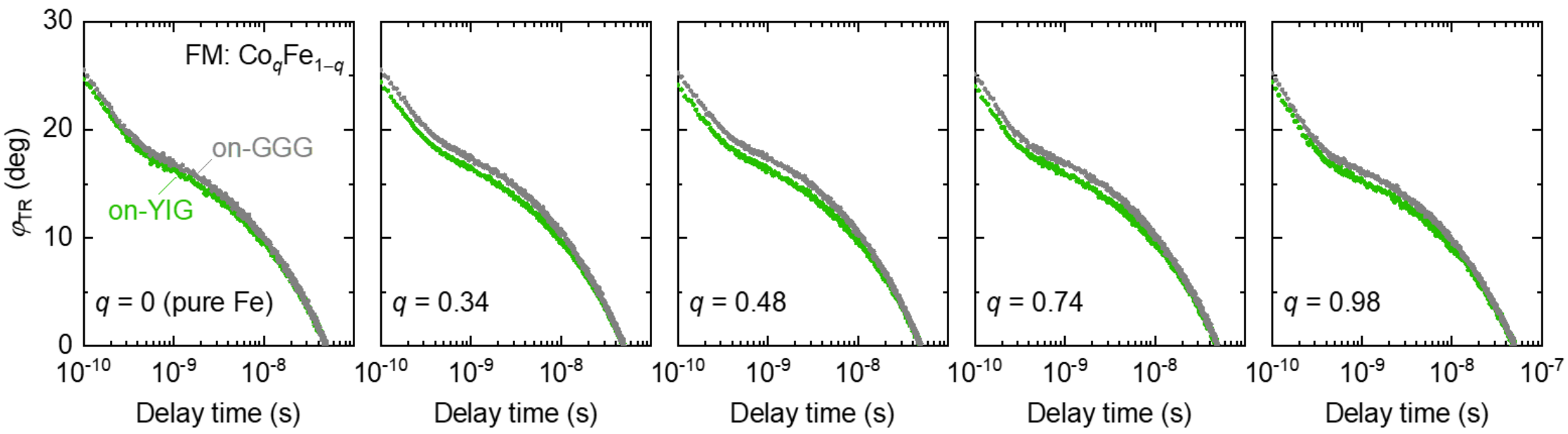

**Figure S9.** Temporal response of $\varphi_{TR}$ for CoFe with the composition ratio $q$ = 0, 0.34, 0.48, 0.74, and 0.98 on GGG and YIG.

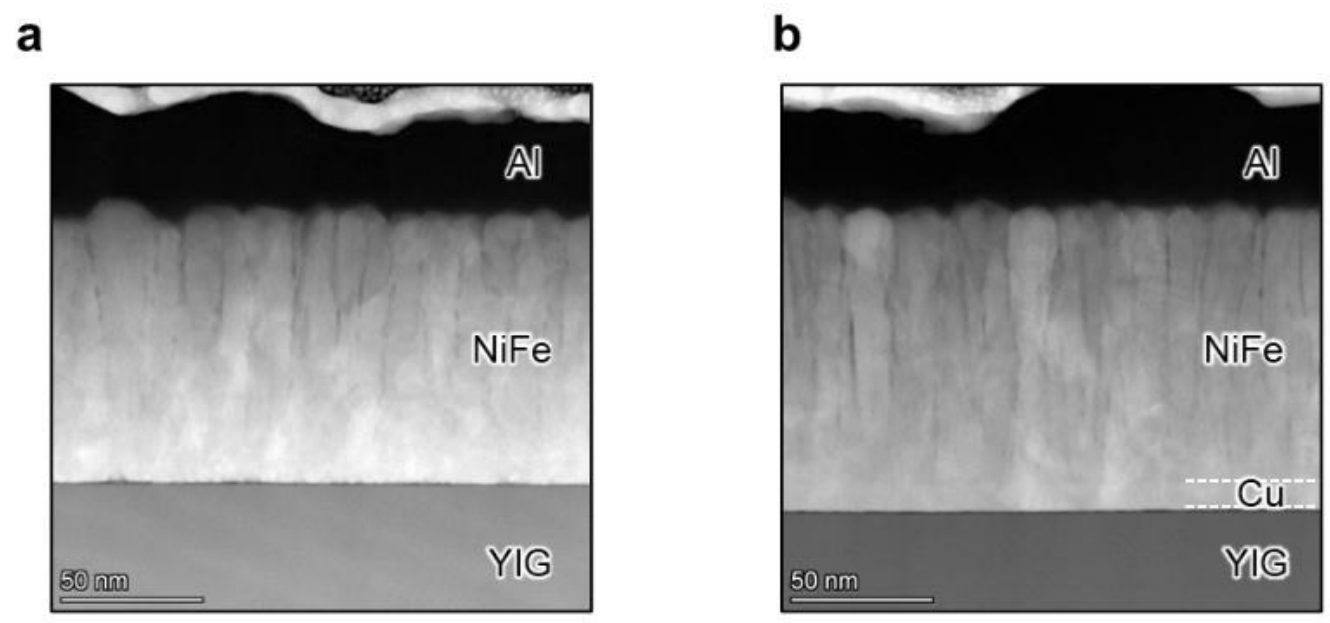

**Figure S10.** HAADF-STEM images of $Ni_{80}Fe_{20}$ film without (a) and with (b) Cu insertion layer on YIG/GGG substrate.

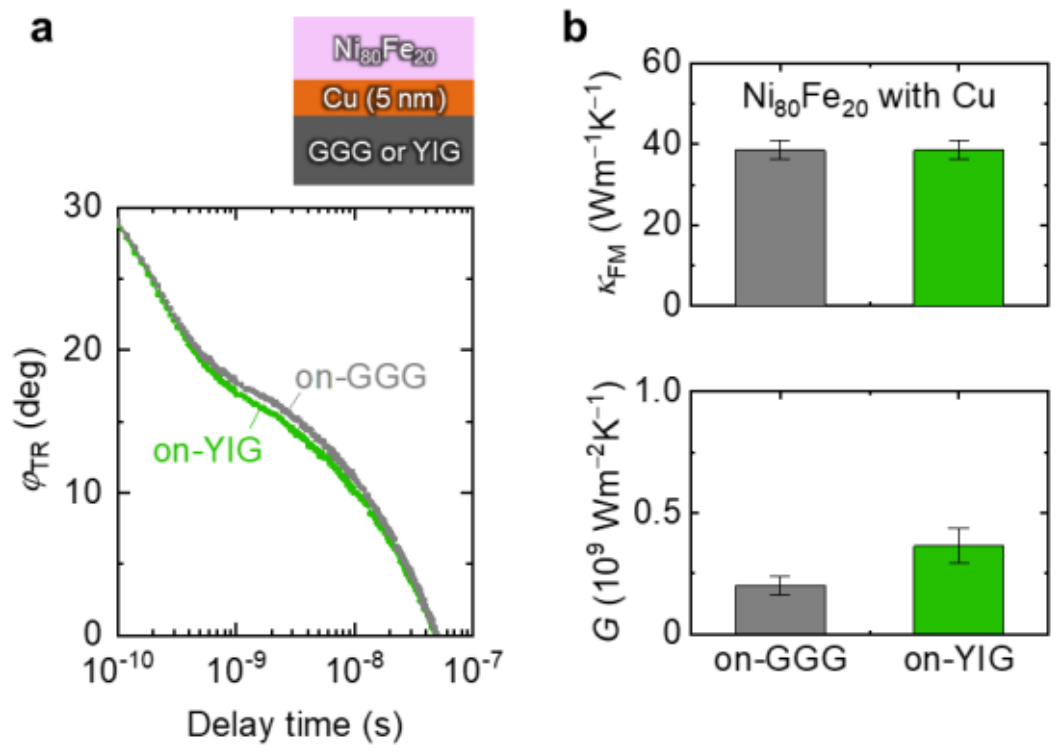

**Figure S11.** a) Temporal response of $\varphi_{TR}$ for $Ni_{80}Fe_{20}$/Cu film with a Cu thickness of 5 nm on GGG and YIG. b) $\kappa_{FM}$ (upper panel) and $G$ (lower panel).

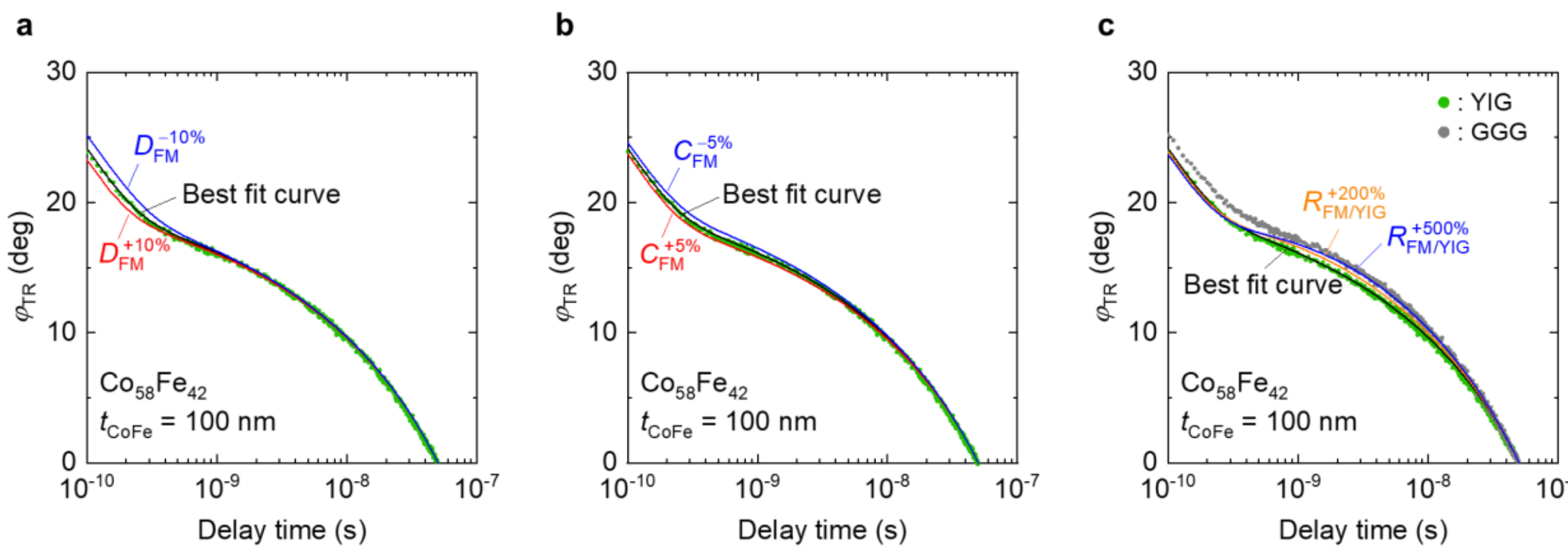

**Figure S12.** Confirmation of measurement accuracy of TDTR for $\kappa_{FM}$ and $R$. Fitting of $\varphi_{TR}$ for $Co_{58}Fe_{42}$ with $t_{FM}$ = 100 nm on YIG with ±10% bounds on the fitted value for thermal diffusivity of CoFe $D_{FM}$ (a), with ±5% bounds for volumetric heat capacity of CoFe $C_{FM}$ (b), and with curves for +200% and +500% values of the fitted value of $R$ (c).

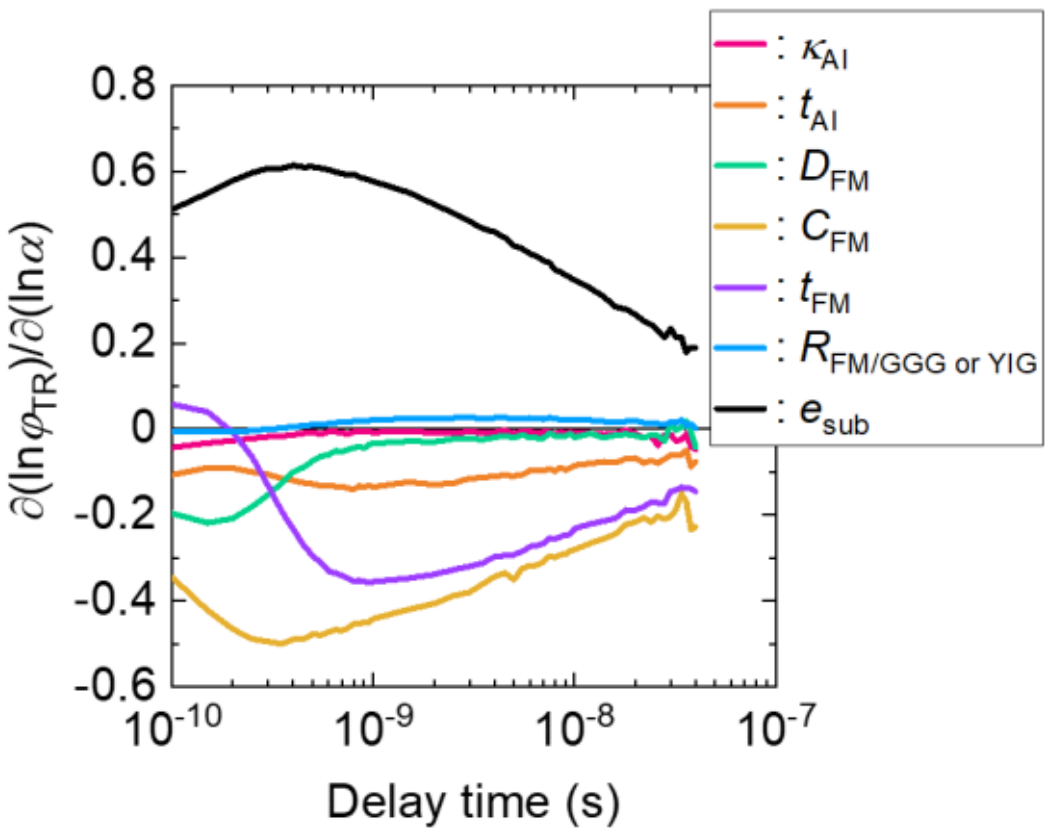

**Figure S13.** Sensitivity of $\varphi_{TR}$ to various parameters $\alpha$ in our thermal model as a function of the delay time for the CoFe film with $t_{FM}$ = 100 nm on YIG or GGG, where $\kappa_{Al}$ is thermal conductivity of Al and $e_{sub}$ thermal effusivity of YIG or GGG.

**Table S1.** Electron spin diffusion length for various ferromagnetic metals at room temperature. Column of Method shows the measurement methods for the electron spin diffusion length of FM $\lambda_{FM}^{e}$, where CPP-GMR presents the current-perpendicular-to-plane giant magnetoresistance, LSV the lateral spin-valve, and TR-MOKE the time-resolved magneto-optical Kerr effect.

| Material | $\lambda_{FM}^{e}$ (nm) | Method | Ref. |
|---|---|---|---|
| Co | 38 ± 12 | CPP-GMR | [S5] |
| Co | 7.7 ± 2 | LSV | [S6] |
| Co | 4.6 ± 1 | TR-MOKE (~10 MHz) | [S7] |
| Fe | 7 ± 1 | TR-MOKE (~10 MHz) | [S7] |
| $Co_{50}Fe_{50}$ | 15 | CPP-GMR | [S8] |
| $Co_{60}Fe_{40}$ | 6.2 | LSV | [S6] |
| $Ni_{81}Fe_{19}$ | 5.2 ± 2 | LSV | [S6] |
| NiFe (unspecified composition) | 3 | LSV | [S9] |